\documentstyle{article}
\input epsf
\newcommand{\cd}{\partial}
\newcommand{\beq}{\begin{equation}}
\newcommand{\eeq}{\end{equation}}

\title{The Kink Casimir Energy in a Lattice Sine-Gordon Model}
\author{J M Speight\thanks{Present address: Department of Mathematics, University of Texas at Austin, Austin TX 78712, USA} \\
        Department of Mathematical Sciences, \\
        University of Durham, Durham DH1 3LE}
\date{}        

\begin{document}
\maketitle

\begin{abstract}

The Casimir energy of quantum fluctuations about the classical kink 
configuration
is computed numerically in the weak coupling approximation for a recently
proposed lattice sine-Gordon model. This energy depends periodically on the 
kink
position and is found to be approximately sinusoidal.

\noindent
\vspace{0.5cm}
PACS classification numbers: 03.65.Sq, 11.10.Lm, 63.10.+a.

\end{abstract}
 
\section{Introduction}

Classically, a topologically stable field configuration may be regarded as
lying in a potential well in an infinite dimensional configuration space. Two
solutions in different sectors are separated by an infinite potential barrier
by virtue of the topology. Quantum mechanically, a particle cannot sit at the 
bottom of a well: it always possesses a zero point energy dependent on the shape
of the well bottom. The analogous situation applies to fields also.

We wish to find the zero point energy associated with a kink configuration in
a certain lattice sine-Gordon model. The model was chosen because it has an exact
lattice version of the topological lower bound on kink energy, and an explicit
static kink which saturates this bound. The calculation is performed in the weak
coupling approximation,
by performing a Taylor expansion of the potential about the kink configuration,
truncating the expansion at quadratic order. We then find the normal modes of
the truncated system, reducing the problem to that of an infinite set of 
decoupled harmonic oscillators. On quantization, each contributes a zero point
energy. Summing over all oscillators gives an infinite total.

If we similarly approximate the vacuum well-bottom we can derive the zero point
energy associated with the trivial vacuum. This is also infinite. The
``physical'' quantity required is the energy difference between the kink and
the vacuum, since we can always {\em define} the vacuum energy to be zero.
This quantity, analogous to the Casimir energy of Quantum Electrodynamics
\cite{cas}, turns out to be finite and less than the classical energy of the
kink. Since the model possesses only discrete translation symmetry, the normal
mode frequencies, and hence the Casimir energy depend periodically on the kink
position.

Calculation of the normal mode frequencies amounts to finding the 
eigenvalues of an infinite-order, tridiagonal, symmetric matrix. In practice
this is not possible and the system must be truncated symmetrically about the
kink centre, ignoring the large  $|x|$ degrees of freedom. It is possible to
show that the resulting quantum energy correction must be negative, for any
size of truncated system. The correction may be calculated numerically.

\section{The Lattice Sine-Gordon Model}

Consider the lattice sine-Gordon model (LSGM)  defined by the Lagrangian
\cite{RSW},
\begin{equation}
L=\frac{h}{4}\sum_{j\in{\cal Z}}\left[\dot{\psi}^{2}_{j}
   -\frac{4}{\lambda^{2}h^{2}}\sin^{2}\frac{\lambda}{2}(\psi_{j+1}-\psi_{j})
   -\frac{m^{2}}{\lambda^{2}}\sin^{2}\frac{\lambda}{2}(\psi_{j+1}+\psi_{j})
                                \right],
\end{equation}
where $h$ is the lattice spacing, $m$ is a mass parameter, $\lambda$ is a
dimensionless coupling constant and ${\cal Z}$ denotes the set of integers.
 Taking the $h \rightarrow 0$ limit one recovers the standard
sine-Gordon model:
\begin{equation}
\lim_{h \rightarrow 0} L = \frac{1}{4}\int_{-\infty}^{\infty}
      dx\,\left[ \left(\frac{\partial\psi}{\partial t} \right)^{2}
                -\left(\frac{\partial\psi}{\partial x} \right)^{2}
                -\frac{m^{2}}{\lambda^{2}}\sin^{2}\lambda\psi \right],
\end{equation}                
where $x=hj$. The standard sine-Gordon Lagrangian is more usually written in 
terms of $\phi =2\psi$ and differs by a factor of $8$, but the above form is
more convenient.

The LSGM has an infinite set of discrete potential minima,
\begin{equation}
\psi = \frac{n\pi}{\lambda}, 
\end{equation}
where $n \in {\cal Z}$,
and a static kink interpolating between neighbouring minima, which may be
derived using a Bogomol'nyi argument. The energy of a static configuration is
just the potential $V$, and
\begin{eqnarray}
0 &\leq& \frac{h}{4}\sum_{j}\left(\frac{2}{\lambda h}
        \sin\frac{\lambda}{2}(\psi_{j+1}-\psi_{j})-\frac{m}{\lambda}
        \sin\frac{\lambda}{2}(\psi_{j+1}+\psi_{j})\right)^{2}
        \nonumber \\
    &=& V +\frac{m}{2\lambda^{2}}\sum_{j}(\cos\lambda\psi_{j+1}
                                          - \cos\lambda\psi_{j})
                                          \nonumber \\
\Rightarrow V &\geq& \frac{m}{\lambda^{2}} \label{eq:bogbound}
\end{eqnarray}
when kink boundary conditions are imposed. This (the Bogomol'nyi) bound is
saturated if and only if
\begin{equation}
\sin\frac{\lambda}{2}(\psi_{j+1}-\psi_{j})=hm 
\sin\frac{\lambda}{2}(\psi_{j+1}+\psi_{j}).
\label{eq:bog}                                                                                                                                                               
\end{equation}

The first order difference equation, (\ref{eq:bog}), is called the Bogomol'nyi 
equation and, remarkably, has an explicit kink solution:
\begin{equation}
\psi_{j}=\frac{2}{\lambda}\tan^{-1}\left[\left(\frac{2+hm}{2-hm}\right)^{j-b/h}\right].
\label{eq:kink}
\end{equation}
The dimensionless parameter $hm \in (0,2)$ for sensible solutions.
The arbitrary constant $b$ may take any real value --- the kink energy is
not position dependent. The right hand side of (\ref{eq:bogbound}) may be 
interpreted as the classical kink mass, $M$. Small velocity kink dynamics can then be
approximated by geodesic motion of a point particle, mass $M$, on the manifold
generated by $b$ translation, with a natural  induced metric \cite{RSW}.

\section{The Weak Coupling Approximation}

We follow the method outlined in \cite{raj} adapted for the infinite lattice.
We work in natural units, effectively having absorbed a factor $\sqrt{\hbar}$
into $\lambda$ so that, by the standard argument,
the weak coupling criterion $\lambda \ll 1$ yields a semi-classical
approximation.

The momentum conjugate to $\psi_{j}$ is
\begin{equation}
\pi_{j} = \frac{\partial L}{\partial \dot{\psi_{j}}} = \frac{h}{2}\dot{\psi_{j}}
\end{equation}
Thus, the LSGM Hamiltonian is
\begin{equation}
H = \sum_{j}\frac{\pi_{j}^{2}}{h} + V(\psi)
\end{equation}
where
\begin{equation}
V(\psi) = \frac{h}{4}\sum_{j}\left[
         \frac{4}{\lambda^{2}h^{2}}\sin^{2}\frac{\lambda}{2}(\psi_{j+1}-\psi_{j})
        +\frac{m^{2}}{\lambda^{2}}\sin^{2}\frac{\lambda}{2}(\psi_{j+1}+\psi_{j})
                             \right]
\end{equation}
Let $\tilde{\psi}$ be a static configuration giving a local (in configuration
space) minimum of the potential, $V(\psi)$. We treat motion about this stable
configuration in the small $\lambda$ approximation by Taylor expansion of
$V$:
\begin{equation}
V(\psi) = V(\tilde{\psi})+
         \frac{m}{2}\sum_{j,k}W_{jk}
         (\psi_{j}-\tilde{\psi}_{j})(\psi_{k}-\tilde{\psi}_{k})+\cdots
\end{equation}
where
\newpage
\begin{eqnarray}
W_{jk} &=& \frac{1}{m}\left.\frac{\partial^{2}V}{\partial\psi_{j}\partial\psi_{k}}\right|_{\tilde{\psi}} \nonumber \\
       &=& \frac{hm}{4}\left(\delta_{j,k}\left\{\frac{2}{(hm)^{2}}
          \left[\cos\lambda(\tilde{\psi}_{k}-\tilde{\psi}_{k-1})
          +     \cos\lambda(\tilde{\psi}_{k+1}-\tilde{\psi}_{k})\right]\right.\right.
          \nonumber \\
       & & +\left.\frac{1}{2}
          \left[\cos\lambda(\tilde{\psi}_{k}+\tilde{\psi}_{k-1})
          +     \cos\lambda(\tilde{\psi}_{k+1}+\tilde{\psi}_{k})\right]
           \right\} \nonumber \\
       & & +\delta_{j,k-1}\left[-\frac{2}{(hm)^{2}}
           \cos\lambda(\tilde{\psi}_{k}-\tilde{\psi}_{k-1})
          +\frac{1}{2}
           \cos\lambda(\tilde{\psi}_{k}+\tilde{\psi}_{k-1})\right]
           \nonumber \\
       & & \left.+\delta_{j,k+1}\left[-\frac{2}{(hm)^{2}}
           \cos\lambda(\tilde{\psi}_{k+1}-\tilde{\psi}_{k})
          +\frac{1}{2}
           \cos\lambda(\tilde{\psi}_{k+1}+\tilde{\psi}_{k})\right]\right).
\label{eq:Wexp} 
\end{eqnarray}
Note that $W$ is a real, symmetric, tridiagonal matrix. Note also that the
next term in the expansion is
\begin{equation}
\frac{1}{3!}\sum_{i,j,k}\left.\frac{\partial^{3}V}
{\partial\psi_{i}\partial\psi_{j}\partial\psi_{k}}\right|_{\tilde{\psi}}
(\psi_{i}-\tilde{\psi}_{i})(\psi_{j}-\tilde{\psi}_{j})(\psi_{k}-\tilde{\psi}_{k}),
\end{equation}
and that the three derivatives with respect to $\psi$ introduce a factor of
$\lambda^{3}$, leaving an overall factor of $\lambda$ after cancelling the
$1/\lambda^{2}$ of $V(\psi)$. Thus, higher corrections are at least of order
$\lambda$, which is why one can truncate the series in the small $\lambda$
limit.

Owing to the symmetry of the $W$ matrix, there exists an orthogonal
transformation $R$ such that
\begin{equation}
W_{jk} = \sum_{l,m} R_{jl}^{T}U_{lm}R_{mk}
\end{equation}
where $U$ is a diagonal matrix. We may reduce the system to a sequence of decoupled
harmonic oscillators by transforming to the rotated coordinates (normal coordinates)
\begin{equation}
\xi_{j} = \sum_{k}R_{jk}(\psi_{k}-\tilde{\psi}_{k}),
\end{equation}
which have conjugate momenta 
\begin{equation}
\eta_{j} = \sum_{k}R_{jk}\pi_{k}.
\end{equation}
Then,
\begin{equation}
\frac{h}{2}\left(H-V(\tilde{\psi})\right)
=\frac{1}{2}\sum_{j}\left(\eta_{j}^{2}+\frac{hm}{2}\Omega_{j}^{2}\xi_{j}^{2}
                    \right),
\end{equation}
where $\Omega_{j}^{2}$ are the eigenvalues of the $W$ matrix, none of which 
can be negative since $\tilde{\psi}$ locally minimizes the potential.

We now quantize in standard canonical fashion, taking $\tilde{\psi}$ to be
first the vacuum, then the kink located at $x=b$. The vacuum ground state energy is
\begin{equation}
E_{\{0\}}^{\scriptscriptstyle 0} = \frac{m}{\sqrt{2hm}}\sum_{j}\Omega_{j}^{\scriptscriptstyle 0}.
\label{eq:Ezero}
\end{equation}

The kink, by virtue of the zero mode $b$ of (\ref{eq:kink}) lies not in a potential well, 
but in a level-bottomed valley meandering through configuration space. One of the normal 
modes is locally tangential to the valley bottom and consequently has vanishing
corresponding eigenvalue (zero frequency). We shall treat this translation mode, $b$,
classically because in the weak coupling approximation it is much heavier than the
orthogonal modes (mass $m/\lambda^{2}$ compared with $m$).

While the translation orbit of the static kink (\ref{eq:kink}) is an equipotential curve,
neighbouring orbits are not: the potential varies periodically along them. So the
eigenvalues of the $W$ matrix $(\Omega_{j}^{\scriptscriptstyle K})^{2}$ will be $b$
dependent with period $h$. The ground state energy of a kink at $b$ is
\begin{equation}
E_{\{0\}}^{\scriptscriptstyle K}(b) = \frac{m}{\lambda^{2}} + \frac{m}{\sqrt{2hm}}\sum_{j}\Omega_{j}^{\scriptscriptstyle K}(b).
\end{equation}
where the sum may be taken over all eigenvalues since the one we wish to omit
is zero anyway.

There is no reason to expect either $E_{\{0\}}^{\scriptscriptstyle K}(b)$ or 
$E_{\{0\}}^{\scriptscriptstyle 0}$ to converge to a finite sum but one 
would expect finite Casimir energy
\begin{equation}
{\cal E}=E_{\{0\}}^{\scriptscriptstyle K}(b)-E_{\{0\}}^{\scriptscriptstyle 0}-\frac{m}{\lambda^{2}}
\label{eq:QME}
\end{equation}
since the lattice spacing $h$ has effectively cut off the ultra violet
divergence problems of the continuum model. We must still take care
when manipulating the divergent sums of (\ref{eq:QME}) that they are suitably 
regulated before being combined. The method of regulation is determined by
practical considerations: we calculate $E_{\{0\}}^{\scriptscriptstyle K}(b)$ and
$E_{\{0\}}^{\scriptscriptstyle 0}$ on a finite lattice, compute ${\cal E}$ and 
then allow the lattice size to grow large. In practice we must truncate the
lattice to finite size anyway in order to solve the kink matrix eigenvalue
problem.

\section{The Eigenvalue Problem}

We now address the problem of finding the eigenvalues of the vacuum and kink 
$W$ matrices, $W^{\scriptscriptstyle 0}$ and $W^{ \scriptscriptstyle K}$.
Let us first consider the vacuum matrix, obtained from (\ref{eq:Wexp}) by
substituting $\tilde{\psi}_{k} = 0$:
\begin{equation}
W^{\scriptscriptstyle 0}_{jk}
=\frac{hm}{4}\left[\left(\frac{4}{(hm)^{2}}+1\right)\delta_{j,k}
-\left(\frac{2}{(hm)^{2}}-\frac{1}{2}\right)\left(\delta_{j,k-1}+\delta_{j,k+1}
                                            \right)\right].
\end{equation} 
This is a very simple matrix. All the diagonal entries are equal, as are all
the upper and lower diagonal entries. Consequently, the spectrum of the
truncated matrix of order $N$ is known exactly (see, for example, \cite{bell}):
\begin{equation}
(\Omega_{j}^{\scriptscriptstyle 0})^{2}
=\frac{4+h^{2}m^{2}}{4hm}-\frac{4-h^{2}m^{2}}{4hm}\cos\left(\frac{j\pi}{N+1}  
                                                       \right),
\end{equation} 
$j=1,2,\ldots,N$. In the $N\rightarrow\infty$ limit this is strikingly
similar to the dispersion relation for phonons on the lattice \cite{RSW} because the 
normal modes of 
oscillation about the vacuum {\em are} phonons. 

The question of making sense of the $N\rightarrow\infty$ limit of
$\sum_{j}\Omega_{j}^{\scriptscriptstyle 0}$ is irrelevant because there is no
hope of finding the exact spectrum of the kink $W$ matrix, 
$W^{\scriptscriptstyle K}$. Substituting (\ref{eq:kink}) into (\ref{eq:Wexp})
we obtain an explicit expression for $W^{\scriptscriptstyle K}(b)$.
Since the kink is highly localized (provided $hm$ is not too small), at large
$|j|$ the kink configuration rapidly approaches neighbouring vacuum minima
so that $W^{\scriptscriptstyle K}_{jk}(b)$ tend to
$W^{\scriptscriptstyle 0}_{jk}$ away from the matrix centre. The suggestion, then,
is that we truncate the lattice, pinning all the large $|j|$ degrees of
freedom, for which the kink and vacuum configurations are essentially
identical, to their classical values. The resulting finite problem may be
solved numerically.

One can prove that, whatever the order, $N$ (an odd integer), of the 
symmetrically truncated system (that is, truncated symmetrically about the
lattice site $j=0$), the Casimir energy must be 
negative. The proof rests on the observation that, truncated to order
$N=2n+1$,
\begin{equation}
W^{\scriptscriptstyle K}=W^{\scriptscriptstyle 0}+B+C,
\end{equation}
where $B$ is the symmetric, tridiagonal matrix,
\begin{equation}
B=
\left( \begin{array}{ccccccc}
-\beta_{-n} & b_{-n}        & 0             &        &              &              &        \\
b_{-n}      & -\beta_{-n+1} & b_{-n+1}      &        &              &              &         \\
0           & b_{-n+1}      & -\beta_{-n+2} &        &              &              &          \\
            &               &               & \ddots &              &              &           \\
            &               &               &        & -\beta_{n-2} & b_{n-1}      & 0          \\
            &               &               &        & b_{n-1}      & -\beta_{n-1} & b_{n}       \\
            &               &               &        & 0            & b_{n}        & -\beta_{n} 
\end{array} \right),
\end{equation}
\begin{eqnarray}
b_{j} &=& \frac{hm}{8}\left(1-\cos\lambda(\tilde{\psi}_{j}+\tilde{\psi}_{j-1})\right)
         +\frac{1}{2hm}\left(1-\cos\lambda(\tilde{\psi}_{j}-\tilde{\psi}_{j-1})\right) \\
      &\geq& 0, \nonumber   \\
\beta_{j} &=& b_{j}+b_{j+1},
\end{eqnarray} 
and $C$ is the diagonal matrix with elements,
\begin{equation}
C_{ij}=\delta_{ij}\frac{hm}{4}\left[
\cos\lambda(\tilde{\psi}_{j+1}+\tilde{\psi}_{j})+
\cos\lambda(\tilde{\psi}_{j}+\tilde{\psi}_{j-1})-2\right].
\end{equation}
Now, $C$ manifestly has negative semi-definite eigenvalues, 
and $B$ is a special case of a class of matrices whose spectra are known to be
negative semi-definite \cite{WL+GER}. Regarding $B$ and $C$ as 
perturbations to the matrix $W^{\scriptscriptstyle 0}$, we apply a
corollary of the minimax theorem \cite{wilk}:
\begin{quote}
{\bf Theorem:} {\em  If $\Theta,\Upsilon,\Gamma$ are symmetric $N \times N$
matrices with eigenvalues $\theta_{r},\upsilon_{s},\gamma_{t}$ (all sets 
arranged in non-increasing order), and
$$
\Theta=\Upsilon+\Gamma,
$$
then
$$
\theta_{r}\leq \upsilon_{r}+\gamma_{1}
$$
for all $r=1,2, \ldots,N$.}
\end{quote}

Thus the eigenvalues of $W^{\scriptscriptstyle 0}+B$ are shifted down relative
to those of $W^{\scriptscriptstyle 0}$ (the greatest eigenvalue of $B$ is
negative or zero), and similarly, the eigenvalues of
$W^{\scriptscriptstyle 0}+B+C=W^{\scriptscriptstyle K}$ are shifted down
relative to those of $W^{\scriptscriptstyle 0}+B$. Hence,
\begin{equation}
\sum_{j}\Omega^{\scriptscriptstyle K}_{j}(b) \leq 
\sum_{j}\Omega^{\scriptscriptstyle 0}_{j},
\end{equation} 
and, substituting into (\ref{eq:QME}),
\begin{equation}
{\cal E}(b) \leq 0,
\end{equation} 
the quantum mechanical effect must be to {\em lower} the kink energy.
                  
We have assumed, on physical grounds, that
\begin{equation}
{\cal E}_{N}(b) = \frac{m}{\sqrt{2hm}}\sum_{j}\left(\Omega^{\scriptscriptstyle K}_{j}(b)-\Omega^{\scriptscriptstyle 0}\right),
\end{equation}
the Casimir energy of the order $N$ truncated system converges to a constant
value as $N$ grows large. Numerical evidence for this is presented in figure 1,
a graph of $N$ against ${\cal E}_{N}(0)/m$ for various values of $hm$.
Convergence is very fast for large $hm$ (the $hm=1.9$ curve is essentially flat
for $N\geq3$) but much slower for small $hm$.
 Of course, this happens because
the more finely space is discretized, the more degrees of freedom the kink
structure is spread over. There is also a much smaller exacerbating effect due
to the $hm$ dependence in the kink solution, (\ref{eq:kink}) --- large $hm$ kinks
are sharper in ``real'' $x$ space than small $hm$ kinks. Note that, as proved 
above, the Casimir energy is always negative and that its magnitude grows 
large for small $hm$ as we expect from consideration of the
unrenormalized continuum model \cite{dash}. Although $b=0$ was chosen for these 
data, the rates of convergence are virtually independent of $b\in [-h/2,h/2]$.

\vbox{
\centerline{\epsfysize=3truein
\epsfbox[40 170 570 600]{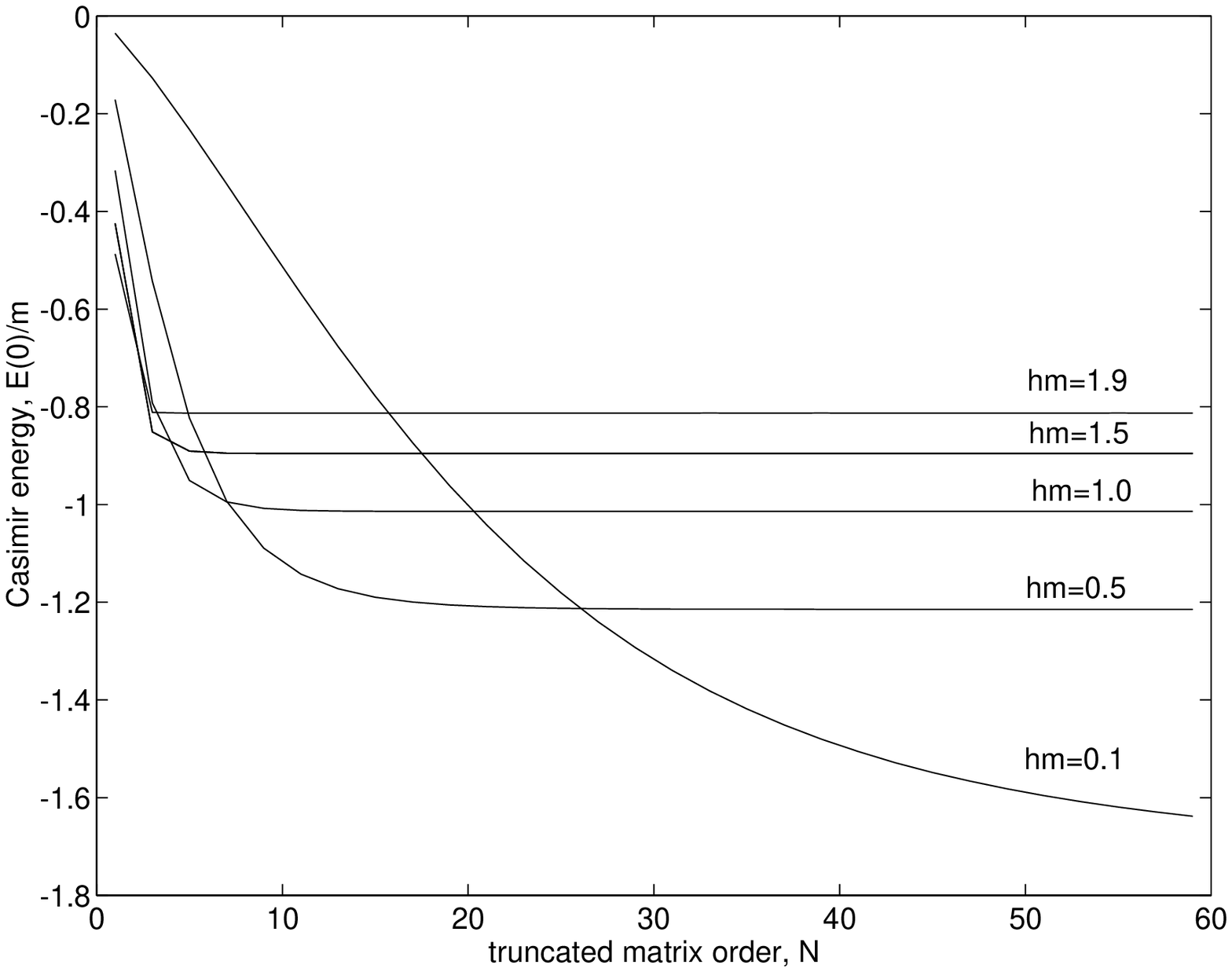}}
\centerline{\it Figure 1: Convergence of the Casimir energy.}
}
\vspace{0.5cm}

\vbox{
\centerline{\epsfysize=3truein
\epsfbox[40 170 570 600]{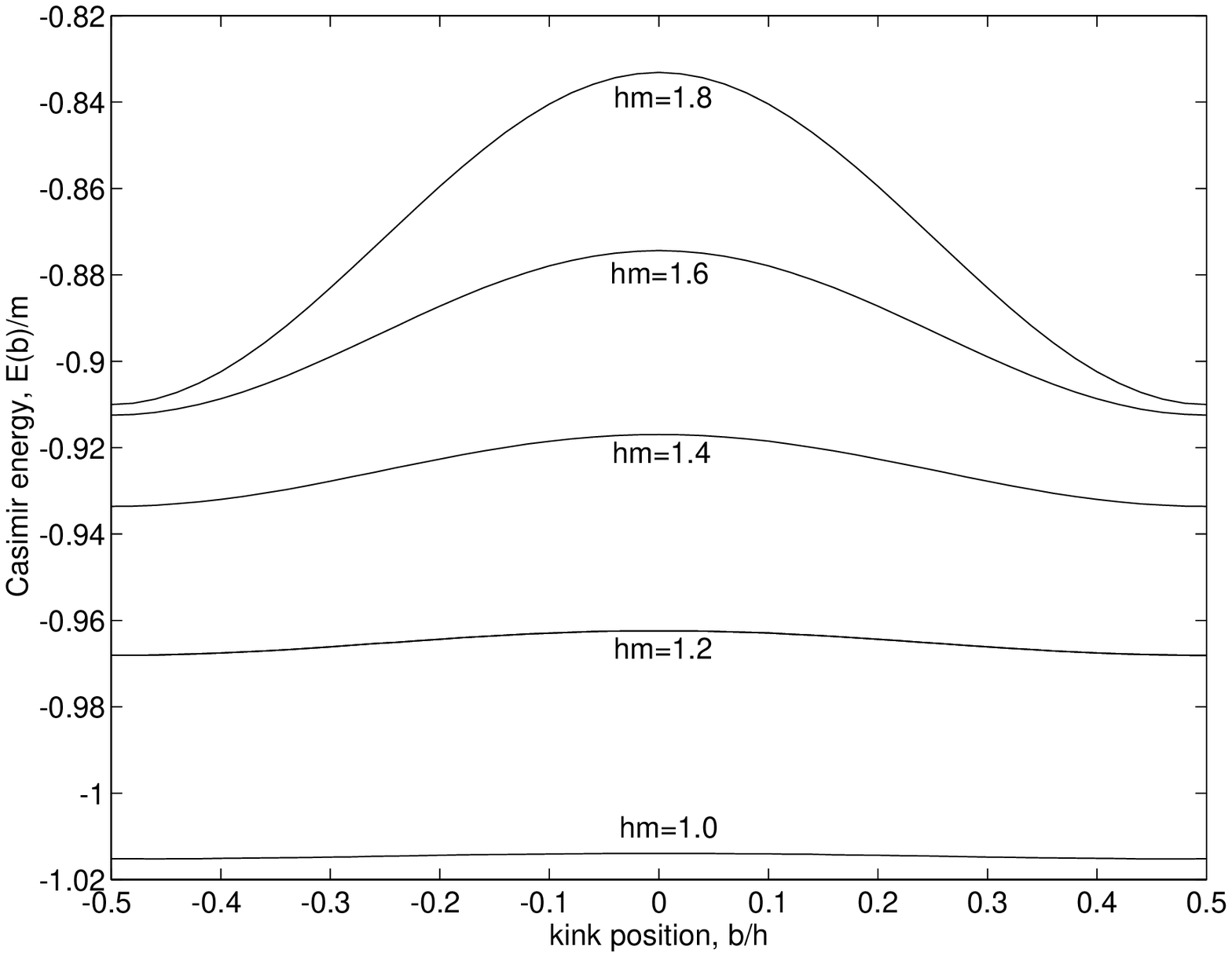}}
\centerline{\it Figure 2: Position dependence of the Casimir energy.}
}
\vspace{0.5cm}

The Casimir energy of suitably truncated systems is plotted as a function of
kink position $b$ in figure 2. It attains its maxima where the kink is
located exactly on a lattice site and its minima where the kink is halfway
between two sites. For large $hm$ the resemblance to a sinusoidal curve is 
remarkable: figure 3 shows a sinusoidal fit for $hm=1.8$. The resemblence 
deteriorates as $hm$ decreases because numerical errors become
relatively large as the Casimir energy barrier $[{\cal E}(0)-{\cal E}(h/2)]/m$
decreases (figure 4). Striking though the similarity is, the author can find no
analytic evidence to support the claim that ${\cal E}(b)$ really {\em is}
sinusoidal.

\vbox{
\centerline{\epsfysize=3truein
\epsfbox[40 170 570 600]{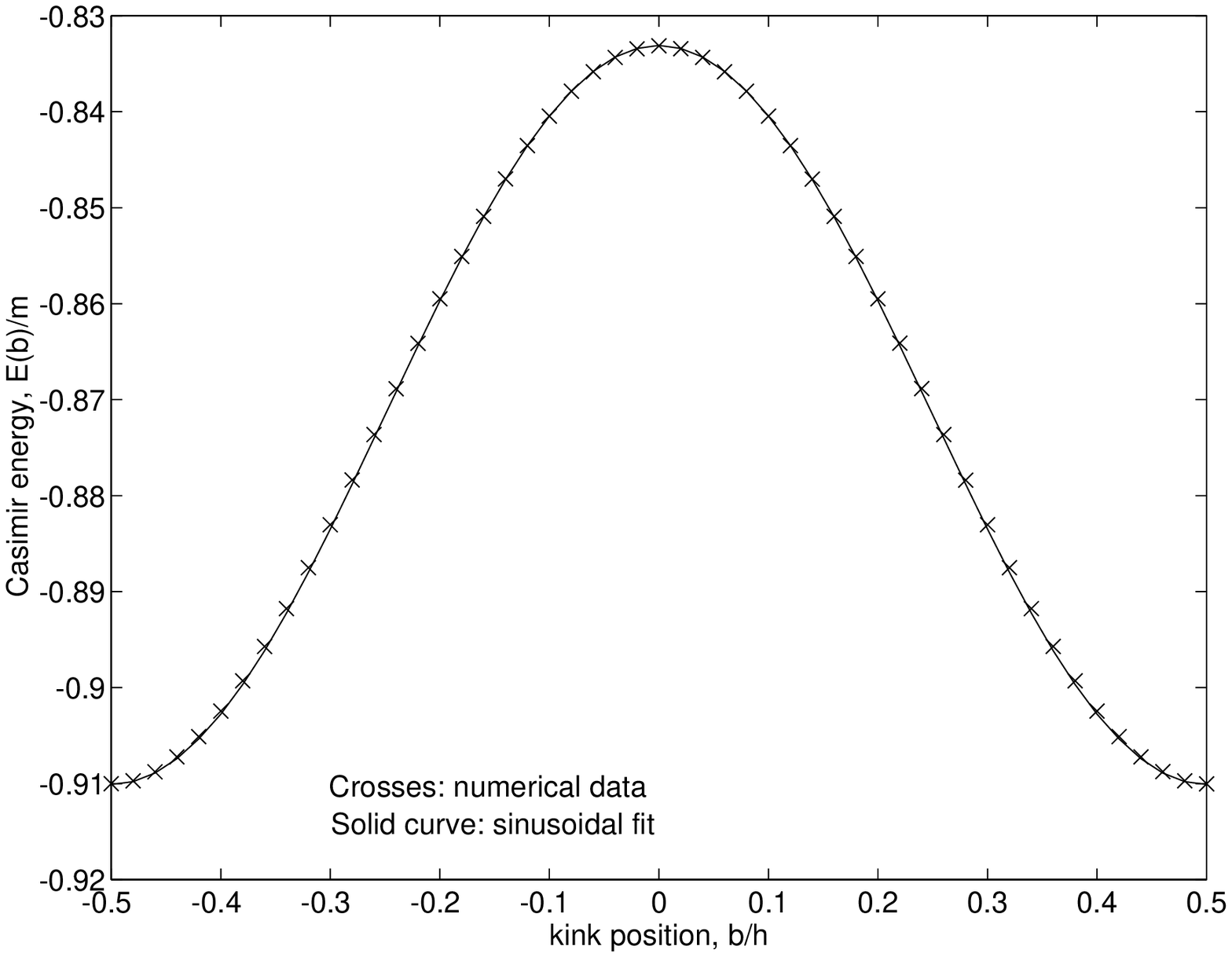}}
\centerline{\it Figure 3: Sinusoidal fit to the $hm=1.8$ Casimir energy.}
}
\vspace{0.5cm}

\vbox{
\centerline{\epsfysize=3truein
\epsfbox[40 170 570 600]{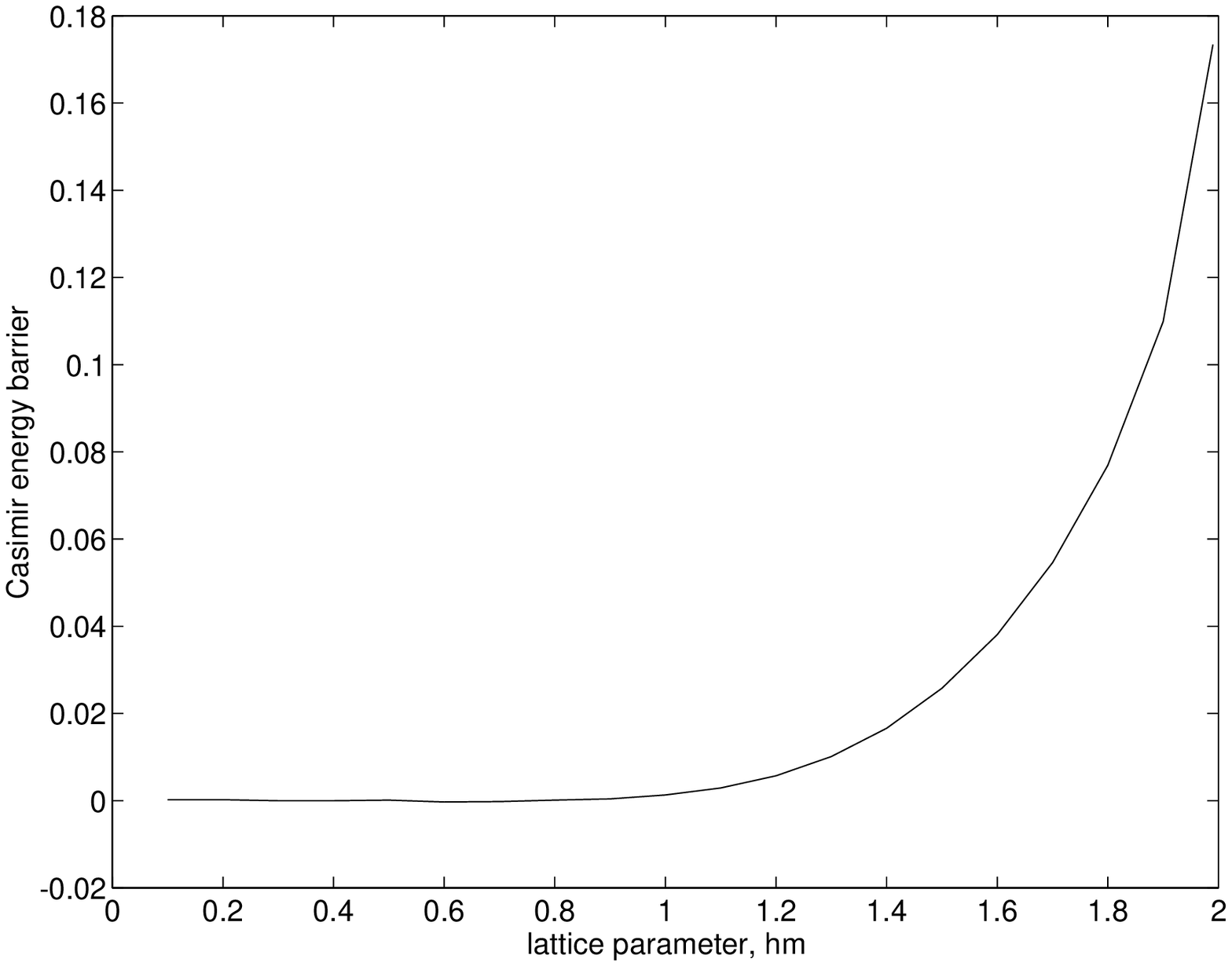}}
\centerline{\it Figure 4: The effect of discreteness on the Casimir energy barrier.}
}
\vspace{0.5cm}

It is the Casimir energy barrier (figure 4) which is physically most relevant
and which will directly affect classical kink dynamics. In the continuum limit
full continuous translation symmetry is recovered, so the barrier should
disappear, as is indicated by the plot. In fact, in common with other (classical
kinetic) discreteness effects of the LSGM \cite{RSW}, the barrier is very small
for all $hm < 1$. If ${\cal E}(b)/m$ for a given $b/h$ is a
strictly increasing function of $hm \in (0,2)$, as it appears to be, then given that ${\cal E}(b)<0$, the barrier should approach a finite value in the 
$hm \rightarrow 2$ limit. Numerical evidence suggests that it does and that
this value is around $0.17$.

\section{Concluding Remarks}

We have seen that quantum fluctuations around the kink configuration spoil
the level kink valley bottom by introducing a Casimir energy which depends
periodically on the (classical) kink position. This energy has been 
computed numerically in the weak coupling approximation and found to be
approximately sinusoidal, maximum when $b=0, \pm h, \pm 2h, \ldots$ and
minimum when $b=\pm h/2, \pm 3h/2, \pm 5h/2, \ldots$, the difference between
these extrema being large for large $hm$ but rapidly vanishing in the
continuum limit. It is superficially similar to the Peierls-Nabarro potential of the Frenkel-Kontorova model \cite{FK} (the conventional lattice sine-Gordon model) but is entirely different in origin, being a purely quantum effect. Since the Casimir effect is a genuine physical phenomenon, experimentally verified in the context of Quantum Electrodynamics, we are led to the conclusion that classical kinks in this lattice model may be ``pinned'' by the quantum mechanics of the orthogonal modes.

Finally, following a suggestion by Gibbons and Manton \cite{mant} we could 
attempt to include the effect of variation of orthogonal mode frequencies
on the quantized geodesic approximation of kink motion by including the
Casimir energy as an extra potential term in the Hamiltonian:
\begin{equation}
\label{eq:ci}
\widehat{H}_{\scriptscriptstyle GA}=-\frac{1}{2}\Delta + {\cal E}(b)
\end{equation}
where $\Delta$ is the covariant Laplacian on the submanifold of static
kinks. Given the periodic nature of ${\cal E}$, we would then expect band
structure in the kink spectrum. There are two objections to this. First, the
kink kinetic terms are of order $\lambda^{2}$ (due to the large kink mass), while the potential was expanded only up to order $\lambda^{0}$, so the
perturbative expansion is not consistent. Second, picking out the $-\frac{1}{2}\Delta$
kinetic term at order $\lambda^{2}$ is also not consistent because there are
several other kinetic terms at this order neglected in equation (\ref{eq:ci}).
This may be seen by re\"{e}xpressing the kinetic Hamiltonian of the full
quantum field theory,
\begin{equation}
\widehat{T}=\frac{1}{h}\sum_{j}\frac{\cd^{2}\,\, }{\cd\psi_{j}^{2}},
\end{equation}
(an infinite dimensional Laplacian) in terms of the kink translation mode $b$
and the orthogonal modes $\xi_{j}$. The resulting formula is very messy, and
includes order $\lambda^{2}$ cross terms with derivatives $\cd^{2}/\cd\xi_{j}
\cd b$ and $\cd^{2}/\cd\xi_{j}\cd\xi_{k}$ in addition to the terms included in
(\ref{eq:ci}). The analogous expression in the continuum model \cite{raj} can
be greatly simplified by boosting to the kink's rest frame, but this trick is not available to us here. These considerations cast doubt on the suggested procedure, at least in
this case. However, one should note that a naive expansion of the Hamiltonian in $\lambda$ may not
be the most physically relevant procedure when considering quantum kink dynamics. For example, one could imagine making the demand that the kink's
``speed'' (in an appropriate quantum sense) be of order $\lambda^{0}$. Given
the kink's large mass in the weak coupling approximation, one would then
expect the kink kinetic term to {\em dominate} the Hamiltonian. So the
legitimacy of the suggested procedure remains an open and somewhat controversial
question.

\vspace{0.5cm}
\noindent
{\bf Acknowledgments:} I would like to thank Bernd Schroers for many useful 
discussions and Richard Ward for advice and comments. I also acknowledge the 
financial support of the UK Science and Engineering Research Council
in the form of a research studentship.

 \end{document}